%% file: W4pPlanar.tex
\documentclass[11pt,a4paper]{article}
\pdfoutput=1

\usepackage{amsmath}
\usepackage[T1]{fontenc}

\usepackage{jheppub}
\usepackage{psfrag}
\usepackage{slashed}
\usepackage{cancel}
\usepackage{lscape}

\usepackage{caption}
\usepackage{array}
\usepackage{graphicx}
\usepackage{subcaption}
\usepackage{multirow}
\usepackage{tabularx}
\usepackage{makecell}

\usepackage[utf8]{inputenc}

\input{ExtraDefinitions.tex}

\newcolumntype{C}[1]{>{\hsize=#1\hsize\centering\arraybackslash}X}%

\newcolumntype{Z}{r<{\hspace{3mm}}}

\title{A numerical evaluation of planar two-loop helicity amplitudes for a W-boson plus four partons}

\author[a]{Heribertus Bayu Hartanto,}
\author[a]{Simon Badger,}
\author[a]{Christian Br\o nnum-Hansen,}
\author[b]{Tiziano Peraro}

\affiliation[a]{Institute for Particle Physics Phenomenology, Department of Physics, Durham University,
Durham DH1 3LE, United Kingdom%
}
\affiliation[b]{Physik-Institut, Universit\"at Z\"urich, Wintherturerstrasse 190, CH-8057 Z\"urich, Switzerland%
}

\emailAdd{
heribertus.b.hartanto@durham.ac.uk,\\
simon.d.badger@durham.ac.uk,\\
christian.bronnum-hansen@durham.ac.uk,\\
peraro@physik.uzh.ch
}

\abstract{
We present the first numerical results for the two-loop helicity amplitudes for the scattering of
four partons and a $W$-boson in QCD. We use a finite field sampling method to reduce directly from
Feynman diagrams to the coefficients of a set of master integrals after applying integration-by-parts
identities. Since the basis of master integrals is not yet fully known analytically, we identify a set of
master integrals with a simple divergence structure using local numerator insertions. This allows for accurate numerical evaluation of the amplitude using sector decomposition methods.
}

\keywords{Perturbative QCD, Scattering Amplitudes}
\preprint{IPPP/19/56, ZU-TH 33/19}

\begin{document}
\maketitle
\flushbottom
\section{Introduction \label{sec:intro}}

The growing precision of high energy collider experiments puts increasing strain on our ability to
make reliable theoretical predictions. Standard techniques for the computation of perturbative
scattering amplitudes often fail when applied to the multi-loop and multi-leg processes currently
produced in abundance at the LHC. Constantly evolving methods have led to differential predictions at
next-to-next-to-leading (NNLO) for $2\to2$ scattering and N${}^3$LO for $2\to1$
processes\footnote{For recent summaries of the state-of-the-art see
\cite{Badger:2016bpw,Bendavid:2018nar}.}.

The need to match experimental precision has led to increasing efforts from the theoretical community
to develop new techniques for $2\to3$ predictions at NNLO. The first hurdle has been to compute
unknown two-loop amplitudes in which the analytic and algebraic complexity causes conventional approaches to
integral reduction to fail. Major new advances that exploit numerical evaluations over finite
fields~\cite{vonManteuffel:2014ixa,Peraro:2016wsq,Peraro:2019svx}
have recently produced the first analytic results for five-parton amplitudes in the leading colour
approximation. Combined with the recently computed analytic master
integrals~\cite{Papadopoulos:2015jft,Gehrmann:2018yef} using the canonical
basis approach to differential equations, a form suitable for combination with the unresolved
contributions to the cross section has been
obtained~\cite{Gehrmann:2015bfy,Badger:2018enw,Abreu:2018zmy,Abreu:2019odu}.

The production of a $W$-boson together with jets at hadron colliders are important signatures that can
be used as precision probes of the Standard Model. QCD corrections to $W$+jets have been a traditional testing
ground for new technology. $pp\to W+j$ was among the first $2\to2$ process computed at NLO
\cite{Giele:1993dj}. The amplitudes for $pp\to W+2j$ were computed using the recently developed on-shell unitarity
method~\cite{Bern:1994zx,Bern:1994cg,Bern:1996ka,Bern:1997sc}
and were implemented into \textsc{MCFM} to provide differential cross-section predictions~\cite{Campbell:2002tg}.
NLO results for associated $W$-boson production with three or more jets are accessible through
automation and the use of the generalised unitarity
method~\cite{Ellis:2008qc,Ellis:2009zw,Berger:2008sj,Berger:2009ep,Berger:2010zx,Bern:2013gka}.

NNLO corrections to $pp\to W+2j$ will open up possibilities for further precision tests of the
Standard Model. The two-loop amplitudes are obvious targets for the new technology developed for massless
five-point amplitudes, yet the off-shell vector boson adds an extra scale and therefore a new layer
of complexity. The first step towards a complete analytic computation is to set up a procedure that
could evaluate the amplitudes numerically using rational kinematics. It is this benchmark evaluation
of the amplitudes that is the subject of this paper.

The computation of higher order corrections to perturbative scattering amplitudes is a well studied
problem. Amplitudes with two or more loops have relied on the technology of
integration-by-parts (IBP)~\cite{Tkachov:1981wb,Chetyrkin:1981qh} reduction, which in recent times
has involved following Laporta's algorithm~\cite{Laporta:2001dd}, together with numerical or
analytic methods for the evaluation of the resulting basis of master integrals. For these
multi-scale basis integrals with massless internal propagators the differential equation
technique~\cite{Kotikov:1990kg,Gehrmann:1999as,Henn:2013pwa}
has been employed to find analytic expressions, most recently for the complete set of
planar~\cite{Gehrmann:2015bfy,Gehrmann:2018yef} and
non-planar integrals~\cite{Abreu:2018rcw,Boehm:2018fpv,Abreu:2018aqd,Chicherin:2018mue,Chicherin:2018old}.
For the case, in which the amplitudes considered here fall, only one of the
three planar families has been
evaluated~\cite{Papadopoulos:2015jft}.
Combining the master integrals into complete amplitudes requires the solution of increasingly
complicated linear systems of IBP equations. Considerable effort has led to a variety of efficient
solutions~\cite{Gluza:2010ws,vonManteuffel:2014ixa,Larsen:2015ped,Ita:2015tya,Kosower:2018obg} and
public
implementations~\cite{Lee:2012cn,Smirnov:2014hma,Smirnov:2019qkx,vonManteuffel:2012np,Maierhoefer:2017hyi}.
Applications to five-particle problems have been possible though yielded large IBP reduction
tables~\cite{Boels:2018nrr,Chawdhry:2018awn}.
In this paper, we only perform the IBP reduction numerically over
finite fields in order obtain the coefficients of the amplitude in
terms of master integrals.  As shown e.g.\ in
refs.~\cite{Badger:2018enw,Peraro:2019svx}, when combined with
functional reconstruction techniques, this approach also allows to
directly reconstruct analytic results for amplitudes, sidestepping the need of
computing and using large analytic IBP tables, which are often
significantly more complicated.

Another important ingredient has been the development of efficient methods to construct
on-shell integrands and integral coefficients. Integrand reduction techniques~\cite{Ossola:2006us}
combined with the use of a Feynman diagram approach or generalised
unitarity have been very successful for the computation of one-loop amplitudes, in particular to construct
scalar integral coefficients numerically. These techniques have been extended to two
loops~\cite{Mastrolia:2011pr,Mastrolia:2012an,Zhang:2012ce,Badger:2012dp,Mastrolia:2012wf,Mastrolia:2013kca,Mastrolia:2016dhn}
and methods to employ unitarity cuts~\cite{Bern:1994zx,Bern:1994cg} to build amplitudes by directly
incorporating IBP decomposition have been
established~\cite{Kosower:2011ty,CaronHuot:2012ab,Ita:2015tya,Abreu:2017xsl,Abreu:2017idw}.

The first steps towards helicity amplitudes for five-point amplitudes were taken through
numerical evaluations of two-loop five point amplitudes in QCD using modular arithmetic~\cite{Badger:2017jhb,
Abreu:2017hqn,Badger:2018gip,Abreu:2018jgq}. These algorithms have been generalised to allow for a full
reconstruction of the coefficients of the pentagon functions classified in~\cite{Gehrmann:2018yef} leading to an analytic form of the
single-minus helicity amplitudes~\cite{Badger:2018enw} and the complete leading colour
five-parton helicity amplitudes within the numerical unitarity framework~\cite{Abreu:2018zmy,Abreu:2019odu}.
The success of computations in the planar sector has shifted focus to the
non-planar sector of massless two-loop five-point amplitudes with a series of new results in
super-symmetric Yang-Mills~\cite{Abreu:2018aqd,Chicherin:2018yne} and
gravity~\cite{Abreu:2019rpt,Chicherin:2019xeg}
as well as in the all-plus sector of QCD~\cite{Badger:2019djh}.

In this paper we consider the case of planar amplitudes with an off-shell external leg.
We apply the recently developed technology for the computation of
two-loop five-particle amplitudes using sampling of Feynman diagrams over finite fields.
Using a modular approach, recently presented as part of the \textsc{FiniteFlow} algorithms~\cite{Peraro:2019svx}, we are able to numerically evaluate the diagrams and perform an integrand reduction, subsequently reducing
the resulting integrals using integration-by-parts identities. Since the complete set
of analytic master integrals is not known, some of the integrals were evaluated numerically
using sector decomposition~\cite{Binoth:2000ps,Smirnov:2015mct,Borowka:2017idc}. Analytic results for the following classes of master integral are available:
one of the three families of the off-shell five-point pentagon-box~\cite{Papadopoulos:2015jft} and four-point functions with
one~\cite{Gehrmann:2000zt} and two off-shell~\cite{Gehrmann:2015ora,vonManteuffel:2015msa,Henn:2014lfa,Caola:2014lpa} legs.
For master integral topologies for which a numerical evaluation through sector decomposition is challenging,
we identified a basis of master integrals using
local numerators~\cite{ArkaniHamed:2010kv,ArkaniHamed:2010gh}
with simplified divergence structure and therefore easier numerical evaluation.
We consider both the $q\bar{Q}Q\bar{q}^\prime\bar\nu\ell$ and $qgg\bar{q}^\prime\bar\nu\ell$ sub-processes
in our computation where the decay of the $W$-boson is also incorporated.

We describe our integrand reduction setup that is subsequently interfaced to IBP reduction in
Section~\ref{sec:setup}. In Section~\ref{sec:w4pamp} we discuss the structure of the leading colour
$W$+4 parton amplitude at two loops including its singularity structure.
The identification of a master integral basis with local numerator insertions is elaborated in
Section~\ref{sec:localMIs}. Finally, we present numerical benchmark results for
both sub-processes
in Section~\ref{sec:results} and draw our conclusions in
Section~\ref{sec:conclusions}.

\section{Calculational framework \label{sec:setup}}

The framework described in this section is a modification of
the numerical algorithm for two-loop amplitudes presented in~\cite{Badger:2018enw} to allow for the use of
integrands built from Feynman diagrams as an alternative to generalised
unitarity cuts in six dimensions. While the unitarity method can be very efficient, a fully
numerical approach, with rational reconstruction, is also able to avoid the traditional problems
associated with the Feynman diagram approach.

We start by generating a set of Feynman diagrams using \textsc{Qgraf}~\cite{Nogueira:1991ex} and
performing colour decomposition to separate the colour parts of the amplitude from the kinematic
parts that depend only on external momenta $\lrbrace{p}$. We obtain
\begin{equation}
\cA^{(2)}_n(\lrbrace{p})  = \sum_{c} \cC_c \; A^{(2)}_{n,c} (\lrbrace{p}),
\label{eq:colourordering}
\end{equation}
where $\cA^{(2)}_n(\lrbrace{p})$ is the two-loop colour-dressed $n$-point amplitude,
$A^{(2)}_n(\lrbrace{p})$ is the two-loop colour-stripped $n$-point amplitude and
$\cC_c$ is the corresponding colour factor. The colour-stripped amplitude is made up of numerator functions,
$N_T(\lrbrace{k},\lrbrace{p})$, and a set of loop propagator denominators,
$D_\alpha(\lrbrace{k},\lrbrace{p})$, for each diagram topology $T$
\begin{equation}
A^{(2)}_n(\lrbrace{p})  = \int \prod_{i=1}^{2} \frac{d^d k_i}{i \pi^{d/2}e^{-\eps \gamma_E}}
\sum_{T}  \frac{N_T(d_s,\lrbrace{k},\lrbrace{p})}{\prod_{\alpha \in T}
D_\alpha(\lrbrace{k},\lrbrace{p})},
\label{eq:numeratorfunction}
\end{equation}
where $k_i$ is the loop momenta, $d=4-2\eps$ is the space-time dimension
and $d_s = g^\mu_\mu$ is the spin dimension.
The loop amplitude in t'Hooft-Veltman (HV) scheme~\cite{tHooft:1972tcz} can be obtained by
setting $d_s = d$, while the Four-Dimensional-Helicity (FDH) scheme~\cite{Bern:2002zk} can be achieved by setting
$d_s=4$. Each numerator function, $N_T(d_s,\lrbrace{k},\lrbrace{p})$, that contains numerators of Feynman
diagrams that share the same diagram topology, is processed by applying
the t'Hooft algebra. This is carried out with the help of \textsc{Form}~\cite{Kuipers:2012rf,Ruijl:2017dtg} and the \textsc{Spinney} library~\cite{Cullen:2010jv}.
In general, the explicit functional dependence of the numerator function at this point is given by
\begin{equation}
N_T(d_s,\lrbrace{k},\lrbrace{p}) =
N_T\big(d_s,k_i.k_j,\mu_{ij},k_i.q_j,q_i.q_j,\bar{u}(p_i)f(k,q)u(p_j)\big),
\label{eq:numeratordep}
\end{equation}
where $q_i = \lrbrace{p_i,\varepsilon_i}$
($\varepsilon_i$ is the polarisation vector of the external
vector boson) and $\bar{u}(p_i)f(k,q)u(p_j)$ is a spinor string made
up of slashed momenta ($\slashed{q}_i$ and $\slashed{k}_i$).
The $d$-dimensional loop momenta can be decomposed into a
four-dimensional part and an extra-dimensional part
\begin{equation}
k_i = \bar{k}_i + \tilde{k}_i\label{eq:loopmomdecomposition}.
\end{equation}
Due to rotational invariance in the extra dimensions, $\tilde{k}_i$ can only appear in the
numerator function as $\mu_{ij} = - \tilde{k}_i \cdot \tilde{k}_j$.

We obtain helicity amplitudes by specifying the helicity/polarisation of each external particles
and we further parametrise the dependence on the external
kinematics by using momentum twistor variables, $x_i$~\cite{Hodges:2009hk}.
This allows us to express the spinor
products of external momenta $(\spA ij, \spB ij)$ and Mandelstam invariants ($s_{ij}$) uniformly in terms
of momentum twistor variables, where momentum conservation and spinor product relations like
Schouten identities are already built in. At this point we are considering the two-loop $n$-point helicity amplitude
\begin{equation}
A^{(2),h}_n(\lrbrace{p})  = \int \prod_{i=1}^{2} \frac{d^d k_i}{i \pi^{d/2}e^{-\eps \gamma_E}}
\sum_{T}  \frac{N^h_T(d_s,\lrbrace{k},\lrbrace{p})}{\prod_{\alpha \in T}
D_\alpha(\lrbrace{k},\lrbrace{p})},
\label{eq:helicitynumerator}
\end{equation}
where the explicit functional dependence on the helicity-dependent numerator function
$N^h_T(\lrbrace{k},\lrbrace{p})$ is
\begin{equation}
N^h_T(d_s,\lrbrace{k},\lrbrace{p}) =
N^h_T\big(d_s,x_i,k_i.k_j,\bar{k}_i.p_j,\mu_{ij},
\spAB {p_a}{\bar{k}_{i}}{p_b}, \spaa {p_a}{\bar{k}_i}{\bar{k}_j}{p_b}, \spbb {p_a}{\bar{k}_i}{\bar{k}_j}{p_b}
\big).
\label{eq:helicitynumeratordep}
\end{equation}
In processing the algebraic expressions in
Eqs.~\eqref{eq:colourordering},~\eqref{eq:numeratorfunction}~and~\eqref{eq:helicitynumerator},
we have used in-house \textsc{Form}~\cite{Kuipers:2012rf,Ruijl:2017dtg} and \textsc{Mathematica}
scripts.

In this form the helicity-dependent numerator functions in Eq.~\eqref{eq:helicitynumerator} must be
re-expressed in terms of integral families that can later be reduced using IBP equations. To achieve
this we apply an integrand reduction algorithm to obtain
\begin{equation}
A^{(2),h}_n(\lrbrace{p})  = \int \prod_{i=1}^{2} \frac{d^d k_i}{i \pi^{d/2}e^{-\eps \gamma_E}}
\sum_{T}  \frac{\Delta^h_T(d_s,\lrbrace{k},\lrbrace{p})}{\prod_{\alpha \in T}
D_\alpha(\lrbrace{k},\lrbrace{p})},
\label{eq:reducedamp}
\end{equation}
where $\Delta$ is the irreducible numerator for an independent topology $T$.
In order to determine $\Delta$, we first need to construct a basis of irreducible
scalar products (ISPs). We opt to use a basis of ISPs in terms of auxiliary propagators
that is suitable for IBP reduction.

To build an IBP compatible integrand basis,
we define an integral family
\begin{align}
G_{a_1 a_2 a_3 a_4 a_5 a_6 a_7 a_8 a_9 a_{10} a_{11}} &=
\int \frac{d^d k_1}{i \pi^{d/2} e^{-\eps \gamma_E}} \frac{d^d k_2}{i \pi^{d/2} e^{-\eps \gamma_E}} \nonumber \\
&\times \frac{1}{k_1^{2a_1}} \frac{1}{(k_1-p_1)^{2a_2}} \frac{1}{(k_1-p_1-p_2)^{2a_3}} \frac{1}{(k_1+p_4+p_{56})^{2a_4}} \nonumber \\
&\times \frac{1}{k_2^{2a_5}} \frac{1}{(k_2-p_{56})^{2a_6}} \frac{1}{(k_2-p_4-p_{56})^{2a_7}} \frac{1}{(k_1+k_2)^{2a_8}} \nonumber \\
&\times \frac{1}{(k_1+p_{56})^{2a_9}} \frac{1}{(k_2+p_1)^{2a_{10}}} \frac{1}{(k_2+p_1+p_2)^{2a_{11}}},
\label{eq:integralfamily}
\end{align}
where $p_{ij \cdots k} = p_i + p_j + \cdots + p_k$.
Up to cyclic permutations of the external legs,
all integrals appearing in Eq.~\eqref{eq:reducedamp} can be written in the form of Eq.~\eqref{eq:integralfamily}.
We follow the conventions used in~\cite{Badger:2018enw} where
negative exponents, $a_i < 0$, correspond to the ISPs of the irreducible numerator, $\Delta_T^h$.
The irreducible numerators are the most general polynomials in the ISPs with exponents bounded by
renormalisability conditions. As an example, the parametrisation for the two-mass double-box topology is
\begin{align}
\Delta^h \bigg(\usepix{1.6cm}{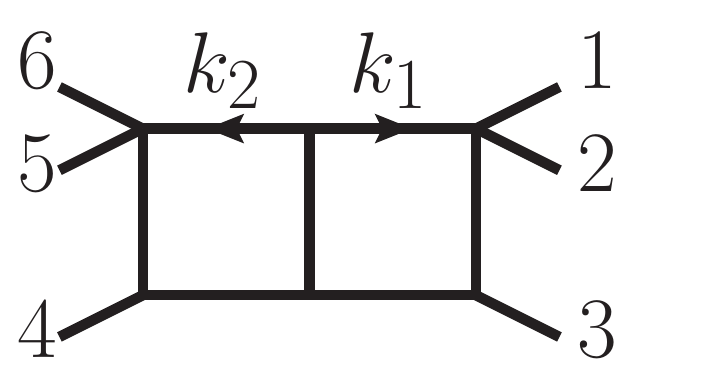}{0}{0}{25}{0}\bigg) =
\sum c^h_{(1,a_2,1,1,1,1,1,1,a_9,a_{10},a_{11})}&(k_1 - p_1)^{-a_2} (k_1+p_{56})^{-2 a_{9}} \nonumber \\
\times &(k_2+p_1)^{-2 a_{10}} (k_2+p_1+p_2)^{-2 a_{11}},\label{eq:twomassdboxbasis}
\end{align}
where the figure represents the topology $T$. The bounds on the exponents are
\begin{alignat}{2}
-4 & \leq a_2 + a_9 &&\leq 0,  \\
-4 & \leq a_{10}+a_{11} &&\leq 0, \\
-6 & \leq a_2 + a_9+a_{10}+a_{10} &&\leq 0.
\end{alignat}
The helicity-dependent coefficients are functions of the spin dimension, $d_s$, and the external kinematics, $c^h = c^h (d_s, \{p\})$. To determine the coefficients we express the numerators in terms of the propagators and ISPs. This is achieved by expanding the loop momenta in terms of external momenta
\begin{align}
\bar{k}_i^\mu &= \sum_{j=1}^{4} a_{ij} p_j^\mu.
\label{eq:loopmomparameterisation1}
\end{align}
The coefficients of the spanning vectors, $p_j$, are functions of the inverse propagators and ISPs, $a_{ij} = a_{ij} (D_\alpha, \text{ISPs})$.
$a_{ij}$ can be determined by solving a linear system of equations constructed by contracting Eq.~\eqref{eq:loopmomparameterisation1}
with the spanning vectors, $p_j$.
All variables in the numerators Eq.~\eqref{eq:helicitynumeratordep} can then be expressed in terms of these coefficients. For example
\begin{align}
\spAB {p_a}{\bar{k}_{i}}{p_b} &= \sum_{j=1}^{4} a_{ij} \spAB {p_a}{p_{j}}{p_b},
\label{eq:loopmomparameterisation2} \\
\mu_{ij} &= - \frac{1}{2} \left( (k_i + k_j)^2 - k_i^2 - k_j^2 \right) + \sum_{m=1}^{4} \sum_{n=1}^{4} a_{im} a_{jn}\, p_m \cdot p_n.
\label{eq:loopmomparameterisation3}
\end{align}
The last relation is obtained by squaring Eq.~\eqref{eq:loopmomdecomposition}.
The variables are straightforwardly evaluated on generalised unitarity cuts by
setting all propagators to zero without relying on explicit loop momenta solutions to the cut constraints.
We observe that this form is the starting point for the derivation of the Baikov
representation~\cite{Baikov:1996iu}, which is obtained by integrating out angular dependence in the
space transverse to the external momenta. We note that the change of variables to rewrite the
numerators in terms of propagators could be performed directly. However, the choice to apply the substitution
using the integrand reduction approach breaks the problem into a series of linear systems with fewer
parameters, rather than one large system.

At this point we can solve for the coefficients of the integrand parametrisations by equating them to the diagram numerators.
Using the two-mass double-box as example again, we have the cut equation
\begin{align}
\Delta^h \bigg(\usepix{1.6cm}{figs/twomassdbox}{0}{0}{25}{0}\bigg) +
\frac{\Delta^h \bigg(\usepix{1.5cm}{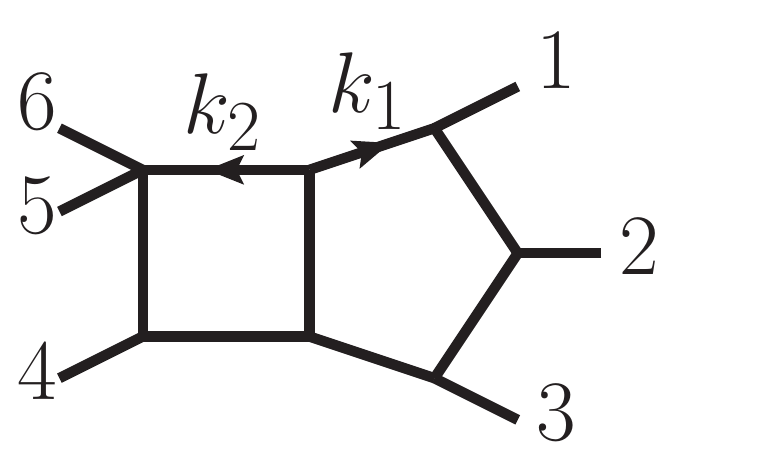}{0}{0}{40}{0}\bigg)}{(k_1-p_1)^2}
 = N^h \bigg(\usepix{1.6cm}{figs/twomassdbox}{0}{0}{25}{0}\bigg)
   + \frac{N^h \bigg(\usepix{1.5cm}{figs/T431_1}{0}{0}{40}{0}\bigg) }{(k_1-p_1)^2},
\end{align}
which is valid only when both sides are evaluated on the hepta-cut
for the two-mass double box, i.e. $D_\alpha = 0$ for $\alpha = 1,3, \dots ,8$.

After setting up the integrand reduction system, we write the helicity amplitude as
a linear combination of integrals in the integral family of Eq.~\eqref{eq:integralfamily},
\begin{equation}
A^{(2),h}_n(\lrbrace{p})  = \sum_{\mathbf{a}} c^h_{\mathbf{a}}(x_i) \;	
G_{\mathbf{a}},
\label{eq:ampreadyforIBP}
\end{equation}
where we sum over tuples $\mathbf{a} = (a_1,a_2,a_3,a_4,a_5,a_6,a_7,a_8,a_9,a_{10},a_{11})$.
At this stage IBP reduction can be applied to the integrals $G_{\mathbf{a}}$.
This basis choice allows for a simple interface to the IBP reduction. The
integrand is never reconstructed analytically but only sampled numerically. The final
results are independent of the particular parametrisation and therefore the exact form of the integrand is not
the main concern here. Nevertheless, alternative representations of the integrand,
e.g.~\cite{Bourjaily:2017wjl}, may lead to improved efficiency.

The integrals with non-zero coefficients after numerical sampling of the integrand are reduced to
a set of master integrals via IBP identities.
The IBP relations are generated in \textsc{Mathematica} using the Laporta
approach~\cite{Laporta:2001dd}  with the aid of \textsc{LiteRed}~\cite{Lee:2012cn},
and solved numerically over finite fields within the \textsc{FiniteFlow}
framework~\cite{Peraro:2019svx}.
We can finally write the helicity amplitudes in the
master integral basis $J_k$
\begin{equation}
A^{(2),h}_n(\lrbrace{p})  = \sum_{k} c^{\mathrm{IBP},h}_k(x_i,\eps)
\; J_k(\lrbrace{p},\eps).
\label{eq:mirepsamp}
\end{equation}

\section{Planar two-loop $W$ plus four parton scattering \label{sec:w4pamp}}

The number of Feynman diagrams contributing to $q\bar{Q}Q\bar{q}^\prime\bar\nu\ell$
and $qgg\bar{q}^\prime\bar\nu\ell$ processes at leading colour are 210 and 603, respectively,
and the leading colour partial amplitudes are extracted according to,
\begin{align}
\mathcal{A}^{(L)}(1_{q},2_{\bar Q},3_{Q},4_{\bar q ^\prime},5_{\bar\nu},6_\ell) &= n^L g_s^2 g_W^2
 \; \;
 \delta_{i_1}^{\;\;\bar i_2} \delta_{i_3}^{\;\;\bar i_4}  \;\;
 A^{(L)}(1_{q},2_{\bar{Q}},3_{Q},4_{\bar q^\prime},5_{\bar\nu},6_\ell),
 \\
\mathcal{A}^{(L)}(1_{q}, 2_{g}, 3_{g}, 4_{\bar q^\prime}, 5_{\bar\nu}, 6_\ell) &= n^L g_s^2 g_W^2
 \; \bigg[
 \left( T^{a_{2}} T^{a_{3}} \right)_{i_1}^{\;\;\bar i_4}
 A^{(L)}(1_{q},2_g, 3_g ,4_{\bar q^\prime}, 5_{\bar\nu}, 6_\ell)
 + (2 \leftrightarrow 3)
 \bigg],
\label{eq:colourdecompositionW}
\end{align}
where $n= m_\eps N_c \alpha_s/(4\pi),\ \alpha_s = g_s^2/(4\pi)$ and $m_\eps=i (4\pi)^{\eps} e^{-\eps\gamma_E}$. $g_s$ and $g_W$ are the strong and weak coupling constants respectively.
We note that the vector boson only couples to the quark line connecting $q$ and $\bar q^\prime$ and
does not couple to the equal flavour quark pair $Q,\bar Q$.

We choose a rational parametrisation of the massless $2\to4$ kinematics using
the momentum twistor parametrisation~\cite{Hodges:2009hk}
\begin{align}
  Z =
  \begin{pmatrix}
    1 & 0 & y_1 & y_2 & y_3 & y_4\\
    0 & 1 & 1   & 1   & 1   & 1 \\
    0 & 0 & 0   & \tfrac{x_5}{x_2} & x_6 & 1 \\
    0 & 0 & 1   & 1   & x_7 & 1-\tfrac{x_8}{x_5}
  \end{pmatrix},
  \label{eq:ZmatrixW}
\end{align}
where $y_i = \sum_{j=1}^{i} \prod_{k=1}^{j} \frac{1}{x_k}$ and
\begin{align}
  x_1 &= s_{12}, &
  x_2 &= -\frac{\spA23\spA41}{\spA12\spA34}, &
  x_3 &= -\frac{\spA34\spA51}{\spA13\spA45}, &
  x_4 &= -\frac{\spA45\spA61}{\spA14\spA56}, \nonumber\\
  x_5 &= \frac{s_{23}}{s_{12}}, &
  x_6 &= -\frac{\spAB{5}{3+4}{2}}{\spA51\spB12}, &
  x_7 &= \frac{\spAA{5}{(2+3+4)(2+3)}{1}}{\spA51 s_{23}}, &
  x_8 &= \frac{s_{123}}{s_{12}}. &
\end{align}
We stress that while we generate a complete parametrisation for the $2\to4$
scattering process, analytic expressions could be obtained with only six independent parameters
since the decay of the $W$ boson completely factorises. Since it is easy to generate a rational
parametrisation for $n$-particle scattering of massless particles, it is simplest to start from a
configuration including the decay of the $W$ boson.

\begin{figure}[h]
  \begin{center}
    \includegraphics[width=0.85\textwidth]{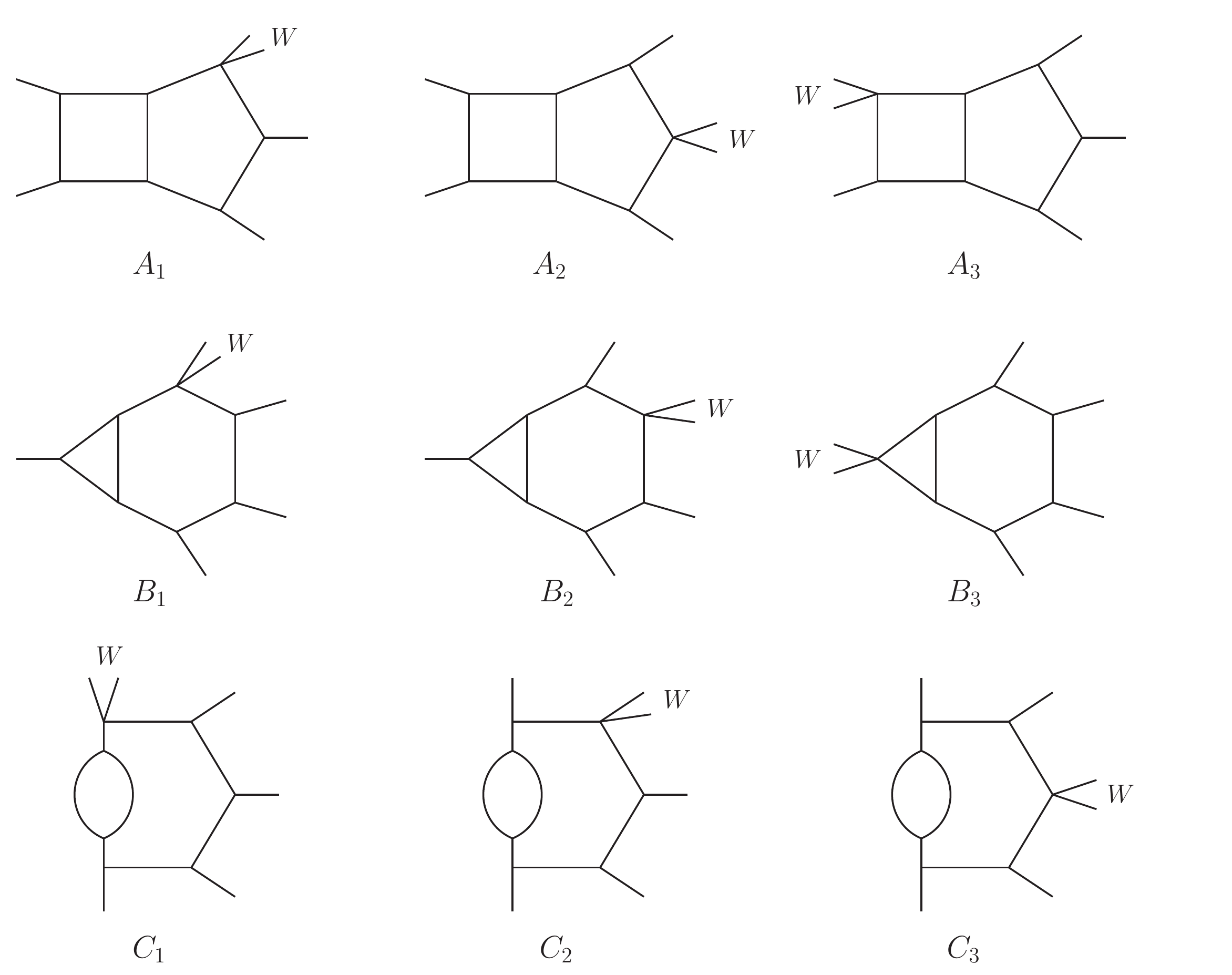}
  \end{center}
  \caption{Independent maximal cut topologies contributing to planar $W+4$ parton scattering at
  two-loops. The full set of 15 maximal cuts can be obtained by including 2 permutations of
  $A_1$, $A_3$, $B_1$, $B_2$, $C_1$ and $C_2$ topologies.}
  \label{fig:W4pmaxcuts}
\end{figure}

\begin{figure}[h]
  \begin{center}
  \includegraphics[width=0.8\textwidth]{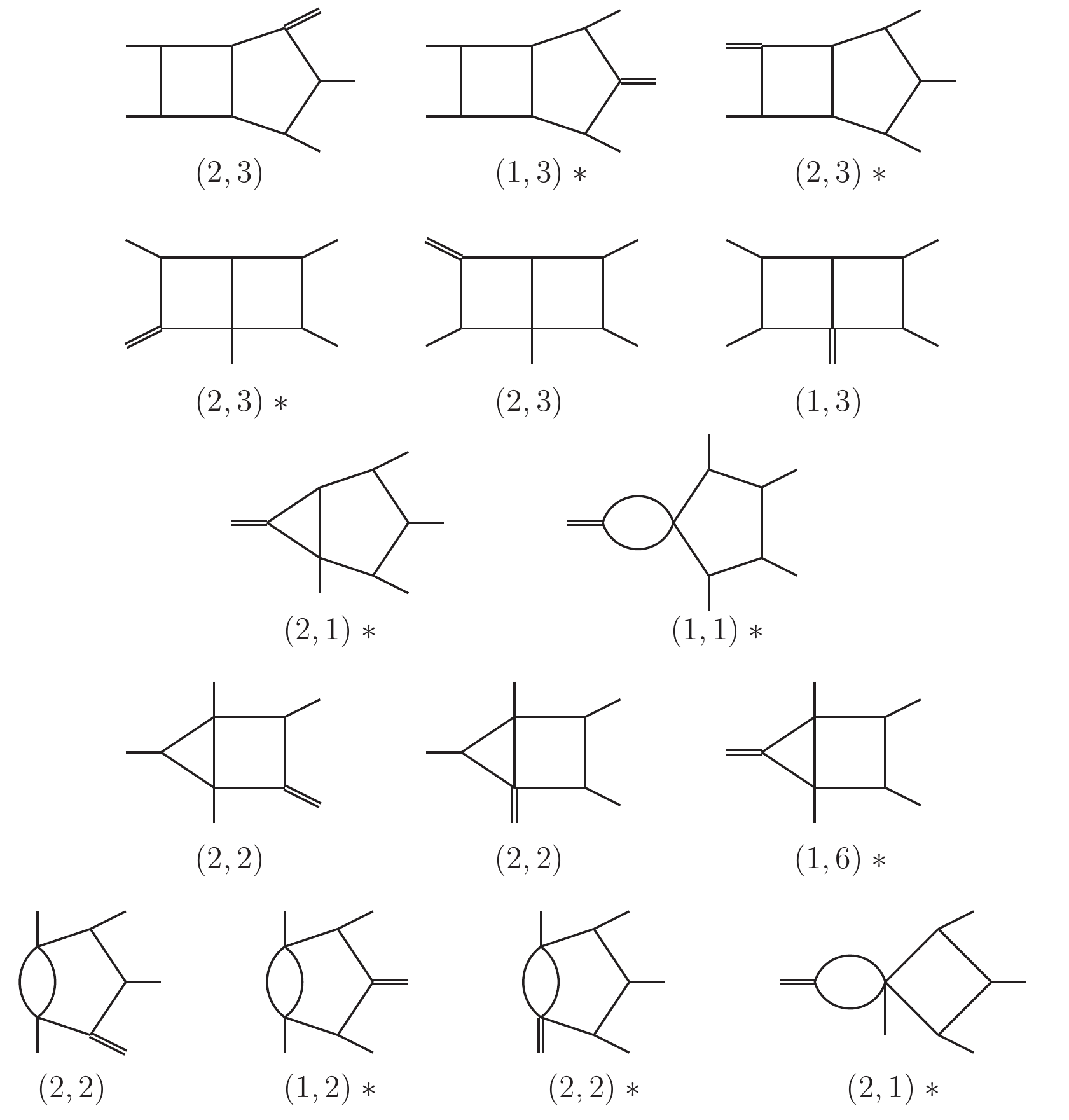}%
  \end{center}
  \caption{Master integrals for leading colour $W+4$ parton scattering at two loops with five external legs. $(a,b)$
  represents the number of crossing of external legs ($a$) and the number master integral for a given
  topology ($b$). A massless (massive) external leg is indicated by a single (double) line external
  leg. The $\ast$ sign identifies master integral topologies that are not known analytically.}
  \label{fig:W4pMasters5pt}
\end{figure}

\begin{figure}[h]
  \begin{center}
  \includegraphics[width=0.8\textwidth]{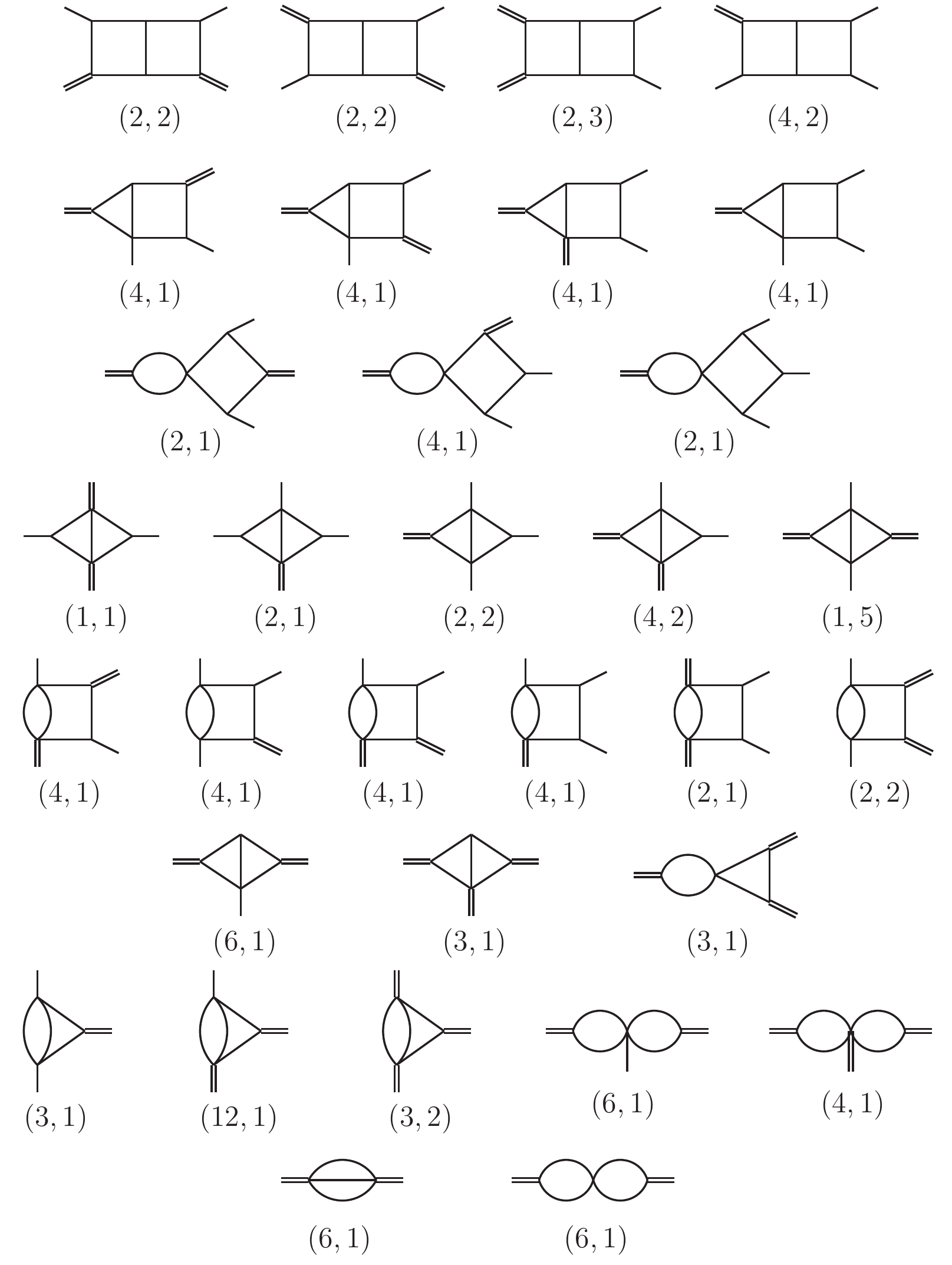}%
  \end{center}
  \caption{Master integrals for leading colour $W+4$ parton scattering at two loops with four external legs
  or fewer. $(a,b)$ represents the number of crossing of external legs ($a$) and the number master integral
  for a given topology ($b$). A massless (massive) external leg is indicated by a single (double) line external leg.
  All master integral topologies shown are known analytically.}
  \label{fig:W4pMastersNon5pt}
\end{figure}

The leading colour partial amplitude is passed through an integrand reduction stage which projects
onto a basis of 453 topologies with irreducible numerators written into the basis of the 15 maximal cuts
shown in Figure~\ref{fig:W4pmaxcuts}. The remaining integrals
are then passed through a Laporta style IBP reduction to find a basis of 202 master integrals
(including the 5 cyclic permutations).
The distinct master integral topologies are shown in
Figures~\ref{fig:W4pMasters5pt}~and~\ref{fig:W4pMastersNon5pt}.
The most complicated integrals that need to be reduced are
rank 5 pentagon-boxes, e.g. $G_{11111111-3-1-1}$ according to the notation defined in
Eq.~\eqref{eq:integralfamily}.

Once the amplitude is decomposed in terms of master integrals and the evaluations of master integrals
are available (either analytically or numerically),
we can perform a Laurent expansion in the dimensional regularisation parameter, $\eps$.
The $\eps$-expanded partial amplitude contains a divergent part, manifested
by the poles in $\eps$, and a finite part.
The infra-red (IR) divergent part of the partial amplitude,
obtained after removing the ultra-violet (UV) divergences by introducing a set of counter-terms,
is universally known~\cite{Catani:1998bh,Becher:2009qa,Becher:2009cu,Gardi:2009qi}.

The pole structure of the unrenormalised amplitude in the HV scheme at one and two loops is given by
\begin{align}
\pole^{(1)} & = 2 I_1(\eps) + \frac{b_0}{\eps},  \\
\pole^{(2)} & =   2 I_1(\eps) \bigg( \hat{A}^{(1)} - \frac{b_0}{\eps} \bigg) + 4 I_2(\eps)
                + \frac{2b_0}{\eps} \hat{A}^{(1)} - \frac{b_0^2}{\eps^2} + \frac{b_1}{2\eps},
\label{eq:poles}
\end{align}
where $\hat{A}^{(1)}$ is the unrenormalised one-loop amplitude normalised to the tree-level amplitude.
The $I_2(\eps)$ operator is defined by
\begin{equation}
I_2(\eps) =   - \frac{1}{2}I_1(\eps) \left[ I_1(\eps)
              + \frac{\beta_0}{\eps} \right]
              + \frac{N(\eps)}{N(2\eps)} \left[ \frac{\beta_0}{2\eps}
                        + \frac{\gamma_{1}^{\cusp}}{8} \right] I_1(2\eps)
              + H^{(2)}(\eps),
\end{equation}
while the $I_1(\eps)$ operators for the $W+4$ parton process at leading colour are
\begin{align}
I^{q\bar{Q}Q\bar{q}^\prime\bar\nu\ell}_1(\eps) &= -N_c \frac{N(\eps)}{2} \bigg( \frac{1}{\eps^2} + \frac{3}{2\eps} \bigg) \big[ \left(-s_{12}\right)^{-\eps} + \left(-s_{34}\right)^{-\eps} \big], \\
I^{qgg\bar{q}^\prime\bar\nu\ell}_1(\eps) &= -N_c \frac{N(\eps)}{2} \bigg\lbrace \bigg( \frac{1}{\eps^2} + \frac{5}{3\eps} \bigg) \big[ \left(-s_{12}\right)^{-\eps} + \left(-s_{34}\right)^{-\eps} \big]
                 + \bigg( \frac{1}{\eps^2} + \frac{11}{6\eps} \bigg) \left(-s_{23}\right)^{-\eps} \bigg\rbrace,
\end{align}
where $N(\eps) = {e^{\eps \gamma_E}}/{\Gamma(1-\eps)}$ and
\begin{align}
H_{q\bar{Q}Q\bar{q}^\prime\bar\nu\ell}^{(2)}(\eps) &= \frac{1}{16\eps} \bigg\lbrace 4 \gamma_1^q - \gamma_1^\cusp \gamma_0^q + \frac{\pi^2}{4} b_0 \gamma_0^\cusp C_F \bigg\rbrace, \\
H_{qgg\bar{q}^\prime\bar\nu\ell}^{(2)}(\eps) &=  \frac{1}{16\eps} \bigg\lbrace 2\left(\gamma_1^q+\gamma_1^g\right)
                                                - \frac{1}{2}\gamma_1^\cusp \left(\gamma_0^q+\gamma_0^g\right)
                                                + \frac{\pi^2}{8} b_0 \gamma_0^\cusp \left(C_F+C_A\right) \bigg\rbrace.
\end{align}
Note that the $H_{q\bar{Q}Q\bar{q}^\prime\bar\nu\ell}^{(2)}(\eps)$ and $H_{qgg\bar{q}^\prime\bar\nu\ell}^{(2)}(\eps)$ functions are given in the leading colour limit.
The $\beta$ function coefficients and anomalous dimensions without the contribution from closed fermion loops $N_f$ are
\begingroup
\allowdisplaybreaks
\begin{align}
\beta_0 = & \;\frac{11}{3}C_A, \\ 
\beta_1 = & \; \frac{34}{3} C_A^2, \\ 
\gamma_0^g = & \; -\frac{11}{3}C_A, \\ 
\gamma_1^g = & \;  C_A^2 \left( -\frac{692}{27} + \frac{11\pi^2}{18} + 2 \zeta_3\right), \\
\gamma_0^q = & \; -3 C_F, \\
\gamma_1^q = & \; C_F^2 \left( -\frac{3}{2} + 2 \pi^2 - 24 \zeta_3 \right)
                  + C_F C_A \left( -\frac{961}{54} -\frac{11\pi^2}{6} + 26 \zeta_3 \right), \\ 
\gamma_0^\cusp = & \; 4, \\
\gamma_1^\cusp = & \; \left( \frac{268}{9} - \frac{4\pi^2}{3} \right) C_A, 
\end{align}
\endgroup
where $C_A = N_c$, $C_F = (N_c^2-1)/(2N_c)$.

With the pole structures available, we can check that the divergent part of the
two-loop amplitude agrees with the predicted UV and IR poles. On the other hand, we can obtain the so-called
two-loop finite remainder by subtracting the UV and IR poles from the two-loop amplitude where
 the pole structure in Eq.~\eqref{eq:poles} is
expanded to $\cO(\eps^0)$. The analytic form
of the finite remainder is in general much simpler than the finite part, as demonstrated
in~\cite{Badger:2018enw,Abreu:2018zmy,Abreu:2019odu,Badger:2019djh}.

\section{Local master integrals \label{sec:localMIs}}

The master integrals appearing in the two-loop leading colour
$q\bar{Q}Q\bar{q}^\prime\bar\nu\ell$ and $qgg\bar{q}^\prime\bar\nu\ell$ amplitudes, that are not known analytically, are evaluated numerically using sector
decomposition~\cite{Binoth:2000ps,Smirnov:2015mct,Borowka:2017idc}.
In general, it is challenging to obtain results with good numerical accuracy for
complicated master integral topologies (e.g. topologies with 6, 7 or 8 propagators in
Figs.~\ref{fig:W4pMasters5pt}~and~\ref{fig:W4pMastersNon5pt}) within
a reasonable amount of time, even for an evaluation in the Euclidean region.
Having numerical accuracy under control is particularly essential when large cancellations occur between
different terms in the amplitude. One way in which this can be achieved is to use
a basis of master integral with local numerator
insertions~\cite{ArkaniHamed:2010kv,ArkaniHamed:2010gh} to regulate divergences. This is the approach
we explore in this work. Another approach well suited to numerical evaluation is to use a quasi-finite basis
of integrals~\cite{Panzer:2014gra,vonManteuffel:2014qoa}. We did not attempt to compare the two
approaches but note that at least two factors are involved: firstly, in the reorganisation of the
amplitude through the change of basis and secondly, the improved convergence of the resulting master
integrals when evaluated with sector decomposition.

As an example, we consider one of the master integral topologies with 8 propagators, $A_1$ pentagon-box topology in Figure~\ref{fig:W4pmaxcuts}.
There are three master integrals for this topology
\begin{equation}
I\bigg(\usepix{1.5cm}{figs/T431_1}{0}{0}{40}{0}\bigg)\big[ 1 \big],\quad
I\bigg(\usepix{1.5cm}{figs/T431_1}{0}{0}{40}{0}\bigg)\big[ (k_1+p_{56})^2 \big],\quad
I\bigg(\usepix{1.5cm}{figs/T431_1}{0}{0}{40}{0}\bigg)\big[ (k_2+p_1)^2 \big],
\label{eq:standardmiA1}
\end{equation}
or
\begin{equation}
G_{11111111000},\quad G_{11111111-100},\quad G_{111111110-10},
\end{equation}
labelled according to Eq.~\eqref{eq:integralfamily}.
In Eq.~\eqref{eq:standardmiA1}, we use a notation for the integral with a numerator insertion
$N(k_i,p_i,\mu_{ij})$
\begin{equation}
 I\big( T \big) \big[ N(k_i,p_i,\mu_{ij}) \big] =
\int \frac{d^d k_1}{i \pi^{d/2} e^{-\eps \gamma_E}} \frac{d^d k_2}{i \pi^{d/2} e^{-\eps \gamma_E}}
\frac{N(k_i,p_i,\mu_{ij})}{\prod_{\alpha \in T} D_\alpha(k_i,p_i)},
\label{eq:integralwnumerator}
\end{equation}
where $T$ is the diagram topology that we will specify by drawing it, and $D_\alpha$ is a set
of massless propagator denominator for a given topology $T$. 
The Laurent expansion for those master integrals starts at
$\cO(\eps^{-4})$ for $G_{11111111000}$ and $\cO(\eps^{-3})$ for $G_{11111111-100}$ and
$G_{111111110-10}$, hence, getting accurate numerical results for the finite part
is a demanding task. Here we follow the notation used in~\cite{Badger:2016ozq} where the local numerators were applied to
six-gluon all-plus helicity amplitudes in Yang-Mills theory.
For the master integral topology under consideration, we choose the following basis of master
integral that exhibits improved IR behaviour
\begin{gather}
 I\bigg(\usepix{1.5cm}{figs/T431_1}{0}{0}{40}{0}\bigg)\big[ \spaa{4}{k_2}{p_{56}}{4} \mu_{11} \big], \nn
 I\bigg(\usepix{1.5cm}{figs/T431_1}{0}{0}{40}{0}\bigg)\big[ \spbb{4}{k_2}{p_{56}}{4} \mu_{11} \big],
 \label{eq:localmisA1} \\
 I\bigg(\usepix{1.5cm}{figs/T431_1}{0}{0}{40}{0}\bigg)\big[ \trm(1(k_1-p_1)(k_1-p_{12})3)
\spaa{4}{k_2}{p_{56}}{4} \big], \nonumber
\end{gather}
where $\tr_{\pm}(ijkl) = \frac{1}{2}\tr(
(1\pm\gamma_5)\slashed{p}_i\slashed{p}_j\slashed{p}_k\slashed{p}_l)$.
The first two integrals in Eq.~\eqref{eq:localmisA1} evaluate to $\cO(\eps)$ and do not contribute to the two-loop
amplitude, while the last integral is finite. Note that, for this topology the master integral
coefficients do not contain any poles in $\eps$, therefore, the master integrals need to be expanded
to $\cO(1)$. To evaluate the last integral in Eq.~\eqref{eq:localmisA1} using sector decomposition
method, we first need to write the numerator insertion in terms of scalar products and momentum
twistor variables ($k_i\cdot k_j$, $k_i\cdot p_j$, $x_i$) using loop momentum decomposition given in
Eqs.~\eqref{eq:loopmomparameterisation1}~-~\eqref{eq:loopmomparameterisation3}
\begin{equation}
 I\bigg(\usepix{1.5cm}{figs/T431_1}{0}{0}{40}{0}\bigg)\big[ \trm(1(k_1-p_1)(k_1-p_{12})3)
\spaa{4}{k_2}{p_{56}}{4} \big] = I\bigg(\usepix{1.5cm}{figs/T431_1}{0}{0}{40}{0}\bigg)\big[
f(k_i\cdot k_j,k_i\cdot p_j,x_i)  \big].
\label{eq:localmiSecDec}
\end{equation}
The integral on the RHS can be directly evaluated using \textsc{pySecDec}~\cite{Borowka:2017idc}
by passing the whole numerator into sector decomposition algorithm. For topologies with four point
kinematics an extra stage of transverse integration is necessary to convert the integrals into a
form compatible with the sector decomposition approach.

\begin{table}
\begin{center}
\begin{tabular}{ m{7em} m{8em} m{3em} m{3em} m{10em}  }
\hline \hline
$T$ & $N(k_i,p_i,\mu_{ij})$ & $\delta_1$ & $\delta_2$ & Crossing \\ \hline\hline
\multirow{3}{7em}{\usepix{1.5cm}{figs/T431_1}{0}{0}{40}{0}}
& $\Omega_{2;4|56}\; \mu_{11}$        & $\eps$ & 1 & \multirow{3}{10em}{ $(1\leftrightarrow
4,2\leftrightarrow 3)$ } \\
& $\Omega_{2;4|56}^*\; \mu_{11}$      & $\eps$ & 1 \\
& $\Psi_{1;1|2|3} \; \Omega_{2;4|56}$ & 1      & 1  \\ \hline
\multirow{3}{7em}{\usepix{1.5cm}{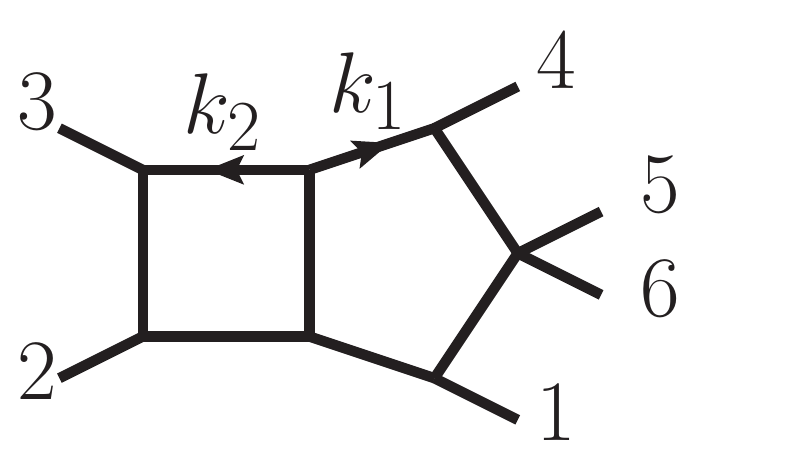}{0}{0}{40}{0}}
& $\Phi_{2;2|3}\; \mu_{11}$        & $\eps$ & 1 & \multirow{3}{10em}{ --- } \\
& $\Phi_{2;2|3}^*\; \mu_{11}$   & $\eps$ & 1 \\
& $\Psi_{1;4|56|1} \; \Phi_{2;2|3}$ & 1      & 1  \\ \hline
\multirow{3}{7em}{\usepix{1.5cm}{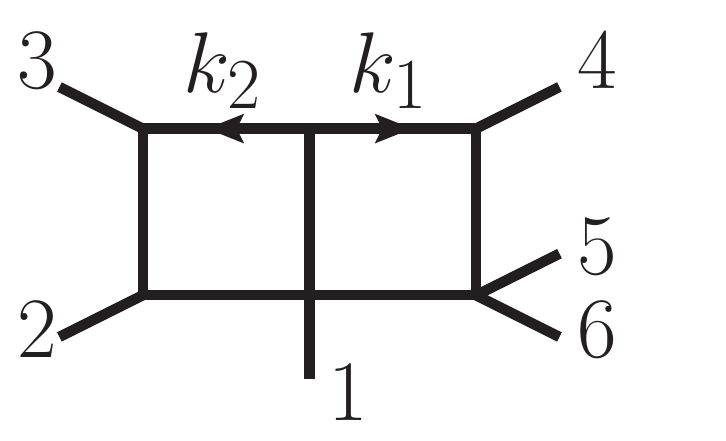}{0}{0}{30}{0}}
& $\mu_{12}$                         & $\eps$ & 1 & \multirow{3}{10em}{ $(1\leftrightarrow
4,2\leftrightarrow 3)$ } \\
& $ \Omega_{1;4|56}\;\Phi_{2;2|3}$   & 1 & 1 \\
& $ \Omega_{1;4|56}\;\Phi_{2;2|3}^*$ & 1 & 1 \\ \hline
\usepix{1.3cm}{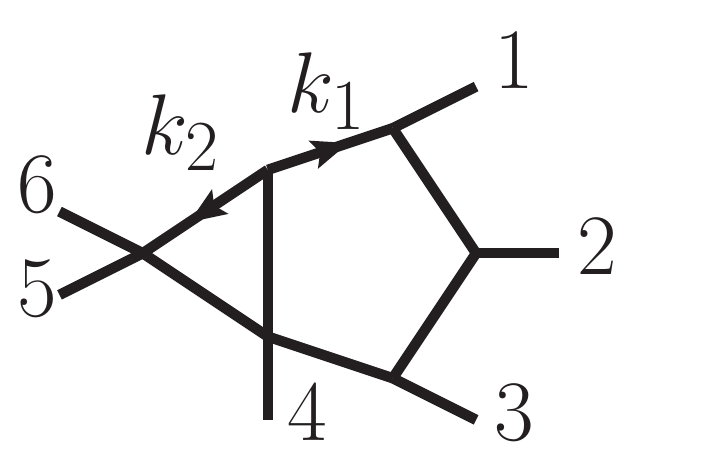}{5}{0}{35}{0} & $\mu_{11}$ & $\eps$ & 1 & $\;(1\leftrightarrow
4,2\leftrightarrow 3)$ \\ \hline
\multirow{6}{7em}{\usepix{1.6cm}{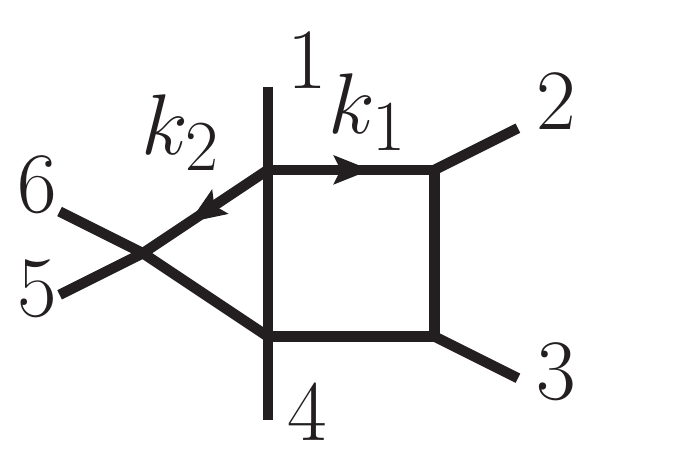}{0}{0}{25}{0}}
& 1 & $\eps^{-2}$ & $\eps$ & \multirow{6}{9em}{ --- } \\
& $\Phi_{1;2|3}$   & 1 & $\eps$ \\
& $\Phi_{1;2|3}^*$ & 1 & $\eps$ \\
& $\Phi_{1;2|3}\;\Omega_{2;1|56}$   & 1 & $\eps$ \\
& $\Phi_{1;2|3}\;\Omega_{2;1|56}^*$ & 1 & $\eps$ \\
& $\Phi_{1;2|3}^*\;\Omega_{2;1|56}$ & 1 & $\eps$ \\ \hline
\multirow{2}{7em}{\usepix{1.1cm}{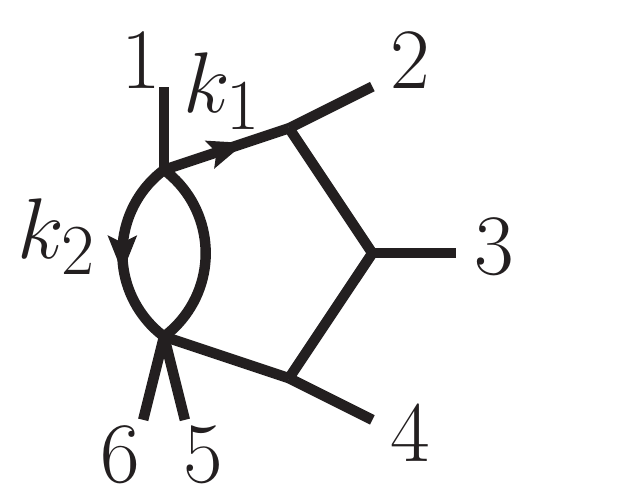}{10}{0}{35}{0}}
& $\Psi_{1;2|3|4}$   & $\eps^{-1}$ & $\eps$ & \multirow{2}{10em}{ $(1\leftrightarrow
4,2\leftrightarrow 3)$  } \\
& $\Psi_{1;2|3|4}^*$ & $\eps^{-1}$ & $\eps$ \\ \hline
\multirow{2}{7em}{\usepix{1.1cm}{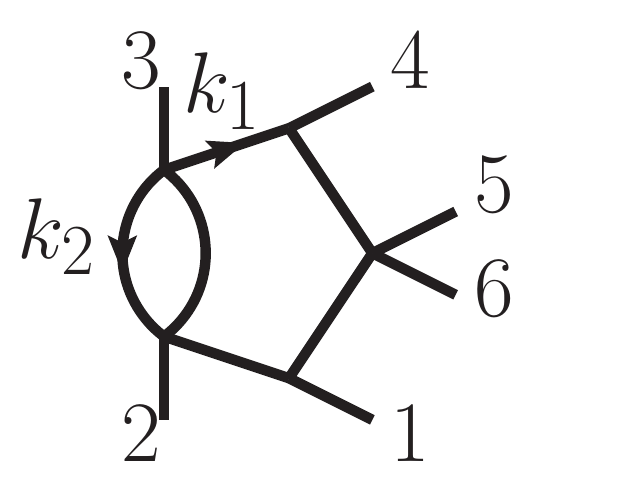}{10}{0}{30}{0}}
& $\Psi_{1;4|56|1}$   & $\eps^{-1}$ & $\eps$ & \multirow{2}{10em}{ ---  } \\
& $\Psi_{1;4|56|1}^*$ & $\eps^{-1}$ & $\eps$ \\ \hline
\end{tabular}
\end{center}
\caption{
Master integral topologies made up of five external legs with local numerator insertions.
The topology $T$ and numerator $N(k_i,p_i,\mu_{ij})$ correspond to the integral definition in
Eq.~\eqref{eq:integralwnumerator}. Numerator building blocks $\Psi$, $\Phi$ and $\Omega$ are defined in
Eq.~\eqref{eq:localnumshorthand}. $\delta_1$ is the order at which the  expansion
of the master integral starts, while $\delta_2$ the highest order in $\eps$ needed from the master integral for
the amplitude evaluation.
}
\label{tab:localMI1}
\end{table}

\begin{table}
\begin{center}
\begin{tabular}{ m{7em} m{8em} m{3em} m{3em} m{10em}  }
\hline \hline
$T$ & $N(k_i,p_i,\mu_{ij})$ & $\delta_1$ & $\delta_2$ & Crossing \\ \hline\hline
\multirow{2}{7em}{\usepix{1.6cm}{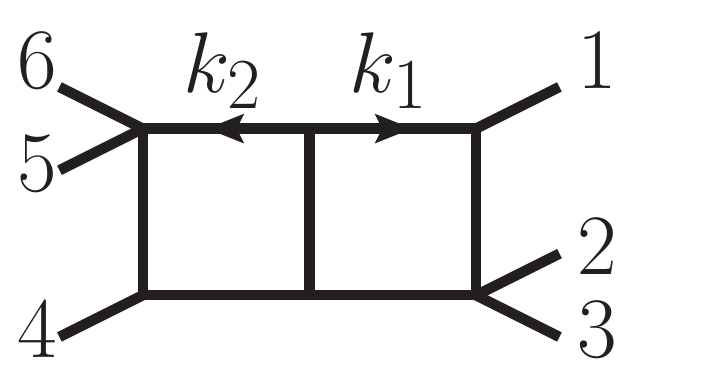}{0}{0}{25}{0}}
& $ \Omega_{1;1|23}\;\Omega_{2;4|56} $     & 1 & 1 & \multirow{2}{10em}{ $(1\leftrightarrow
4,2\leftrightarrow 3)$ } \\
& $ \Omega_{1;1|23}\;\Omega_{2;4|56}^* $   & 1 & 1 \\ \hline
\multirow{2}{7em}{\usepix{1.6cm}{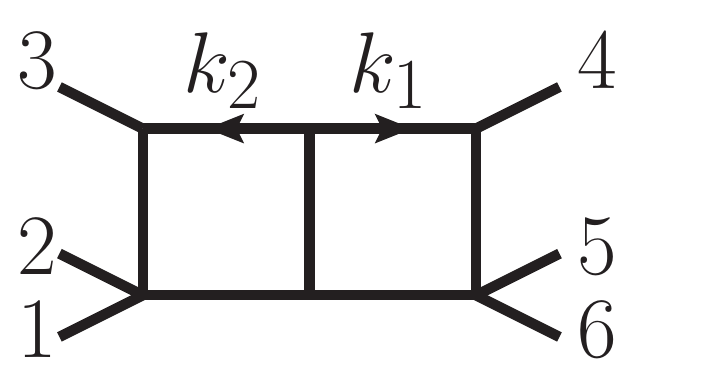}{0}{0}{25}{0}}
& $ \Omega_{1;4|56}\;\Omega_{2;3|12} $     & 1 & 1 & \multirow{2}{10em}{ $(1\leftrightarrow
4,2\leftrightarrow 3)$ } \\
& $ \Omega_{1;4|56}\;\Omega_{2;3|12}^* $   & 1 & 1 \\ \hline
\usepix{1.6cm}{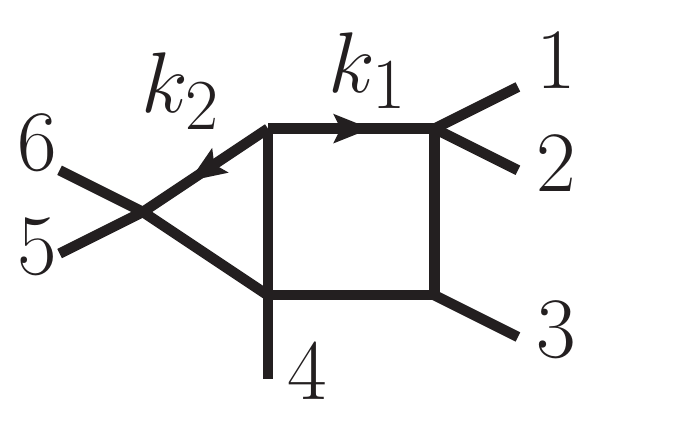}{5}{0}{25}{0} & $ \Omega_{1;3|12} $     & 1 & 1 & $\;(1\leftrightarrow
4,2\leftrightarrow 3)$ \\ \hline
\usepix{1.6cm}{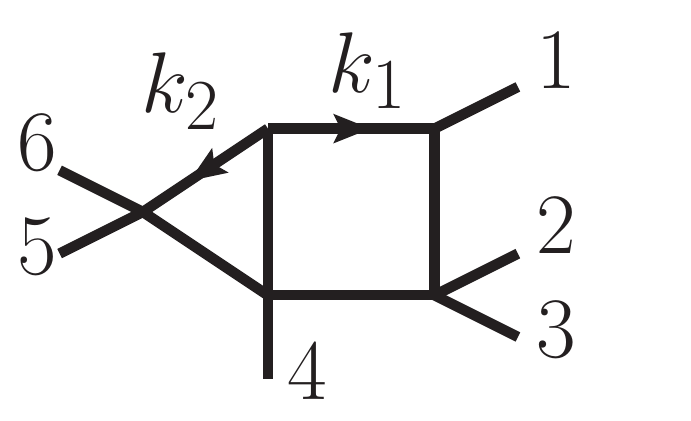}{5}{0}{25}{0} & $ \Omega_{1;1|23} $     & 1 & 1
& \makecell[l]{
  $\;(1\leftrightarrow 4)$ \\
  $\;(2\leftrightarrow 5,3\leftrightarrow 6)$ \\
  $\;(1\leftrightarrow 4,2\leftrightarrow 5,3\leftrightarrow 6)$
  } \\ \hline\hline
\end{tabular}
\end{center}
\caption{
Master integral topologies made up of four external legs with local numerator insertions.
The topology $T$ and numerator $N(k_i,p_i,\mu_{ij})$ correspond to the integral definition in
Eq.~\eqref{eq:integralwnumerator}. Numerator building blocks $\Psi$, $\Phi$ and $\Omega$ are defined in
Eq.~\eqref{eq:localnumshorthand}. $\delta_1$ is the order at which the  expansion
of the master integral starts, while $\delta_2$ the highest order in $\eps$ needed from the master integral for
the amplitude evaluation.
}
\label{tab:localMI2}
\end{table}

In Tables~\ref{tab:localMI1}~and~\ref{tab:localMI2}, we present a list of master integral
topologies with local numerator insertions, where we use
 the following shorthand notation for the numerator insertions
\begin{align}
\Psi_{i;a|b|c}   & = \trm(a (k_i - p_a) (k_i - p_{ab}) c), \nn
\Phi_{i;a|b}     & = \spAB{a}{k_i}{b}, \\
\Omega_{i;a|b} & = \spaa{a}{k_i}{p_b}{a}. \nonumber
\label{eq:localnumshorthand}
\end{align}
We also include in Tables~\ref{tab:localMI1}~and~\ref{tab:localMI2} the order at which the  expansion
of the master integral starts, $\cO(\delta_1)$, the highest order in $\eps$ needed from the master integral for
the amplitude evaluation, $\cO(\delta_2)$, as well as the list of possible permutations/crossings to obtain the
full list of master integral topologies with local numerator insertions.
$\delta_1>\delta_2$ indicates that an integral does not contribute to the finite part of the amplitude, while
$\delta_1=\delta_2$ means that the integral contributes only to the finite part of the
amplitude. In this latter case, it means only the leading order term in the $\eps$ expansion of the
integral is required.

\section{Numerical results \label{sec:results}}

We select a random Euclidean phase-space point by choosing rational values in the momentum twistor
parametrisation from Eq.~\eqref{eq:ZmatrixW},
\begin{gather}
  x_{1} = -1,\quad
  x_{2} = \frac{79}{270},\quad
  x_{3} = \frac{64}{61},\quad
  x_{4} = -\frac{37}{78},\nn \nn
  x_{5} = \frac{83}{102},\quad
  x_{6} = \frac{4723}{9207},\quad
  x_{7} = -\frac{12086}{7451},\quad
  x_{8} = \frac{3226}{2287}.
\label{eq:kinematicpts}
\end{gather}

The numerical results of the leading colour partial amplitudes are obtained by evaluating the master integrals
in three different ways:
\begin{enumerate}
\item We make use of the master integral solutions  that are known analytically
and readily available for evaluation
in the Euclidean kinematics~\cite{Gehrmann:2000zt,Papadopoulos:2015jft}.
\item Master integrals defined with the local numerator insertions,
shown in Tables~\ref{tab:localMI1}~and~\ref{tab:localMI2}, are numerically evaluated using
\textsc{pySecDec}~\cite{Borowka:2017idc}.
\item The remaining master integrals, which contain either one-loop squared
topologies or 5 propagator or fewer two-loop topologies, are evaluated directly using
\textsc{Fiesta}~\cite{Smirnov:2015mct} and
\textsc{pySecDec}~\cite{Borowka:2017idc}.\footnote{The one-loop integrals, required to
$\mathcal{O}(\eps^2)$, are evaluated numerically. Since these terms do not cause any stability
issues there was no need to consider optimisation with analytic expressions.}
\end{enumerate}
Note that there is a class of four-point master integrals with two off shell legs, that are not
covered in~\cite{Papadopoulos:2015jft}, but available
in~\cite{Henn:2014lfa,Gehrmann:2015ora,vonManteuffel:2015msa}, where the solutions are derived
in the physical region. Instead of taking those results and perform analytic continuations
to the Euclidean region for the numerical evaluation, we choose finite local master integral bases
for the 6- and 7-propagator two-loop topologies, as shown in Table~\ref{tab:localMI2}, and
directly evaluate the rest of the integrals in this class numerically\footnote{
These integrals belong to the third type of evaluation discussed above. We see no problem in
performing the analytic continuation on the expressions for the double off-shell $2\to 2$
integrals. However, since the approach with local numerators and the sector decomposition worked with sufficient
accuracy it was unnecessary to do this for our example.}.
To assess uncertainties from the numerical evaluations via the sector decomposition method,
we perform numerical integrations with three different random number seeds.
The final results are obtained by taking the average
and the error is computed by averaging the difference among the three results.

\begin{table}[t]
\begin{center}
\begin{tabularx}{0.95\textwidth}{|C{0.9}|C{0.7}|C{1.1}|C{1.2}|C{1.1}|C{1.0}|}
  \hline
  $qgg\bar{q}^\prime\bar\nu\ell$  & $\eps^{-4}$ & $\eps^{-3}$ & $\eps^{-2}$ & $\eps^{-1}$ & $\eps^{0}$ \\
  \hline
  $\widehat{A}^{(2),[0]}_{-++++-}$ & 4.50000 & -2.38581(9) & 17.0943(2)   & 69.09(3)    & -137.1(3)   \\
  $\widehat{A}^{(2),[1]}_{-++++-}$ & 0.00000 & -0.62498(3) & -147.7288(3) & -347.221(9) & 863.63(8)   \\
  $\widehat{A}^{(2),[2]}_{-++++-}$ & 0.00000 & 0.00000     & -0.031254(7) & -3.72543(2) & -65.7967(5) \\
  \hline
  $qgg\bar{q}^\prime\bar\nu\ell$  & $\eps^{-4}$ & $\eps^{-3}$ & $\eps^{-2}$ & $\eps^{-1}$ & $\eps^{0}$ \\
  \hline
  $\widehat{A}^{(2),[0]}_{-+-++-}$ & 4.50000 & -2.3858(1)   & -16.3282(4)   & -1.397(3)    & 70.05(4) \\
  $\widehat{A}^{(2),[1]}_{-+-++-}$ & 0.00000 & -0.624996(6) & 0.7603(2)     & 4.457(1)     &  1.04(1) \\
  $\widehat{A}^{(2),[2]}_{-+-++-}$ & 0.00000 & 0.00000      & -0.0312498(2) & -0.025800(5) & 0.72620(5) \\
  \hline
  $qgg\bar{q}^\prime\bar\nu\ell$  & $\eps^{-4}$ & $\eps^{-3}$ & $\eps^{-2}$ & $\eps^{-1}$ & $\eps^{0}$ \\
  \hline
  $\widehat{A}^{(2),[0]}_{--+++-}$ & 4.50000 & -2.38579(5)  & -22.18117(7)  & -16.113(6)   & 90.06(4)   \\
  $\widehat{A}^{(2),[1]}_{--+++-}$ & 0.00000 & -0.625000(1) &  1.131987(9)  &  5.7364(2)   & -2.1289(6) \\
  $\widehat{A}^{(2),[2]}_{--+++-}$ & 0.00000 & 0.00000      & -0.0312502(3) &  0.005162(2) & 1.21279(4) \\
  \hline
\end{tabularx}
\end{center}
\caption{The numerical results of
$\hat{A}^{(2),[i]}\left(1_q,2_{g},3_{g},4_{\bar{q}^\prime},5_{\bar\nu},6_\ell \right)$
using kinematic point in Eq.~\eqref{eq:kinematicpts} for each $(d_s-2)$ component.
}
\label{tab:numggDs}
\end{table}

\begin{table}[t]
\begin{center}
\begin{tabularx}{0.95\textwidth}{|C{0.9}|C{0.7}|C{1.1}|C{1.2}|C{1.1}|C{1.0}|}
  \hline
  $q\bar{Q}Q\bar{q}^\prime\bar\nu\ell$  & $\eps^{-4}$ & $\eps^{-3}$ & $\eps^{-2}$ & $\eps^{-1}$ & $\eps^{0}$ \\
  \hline
  $\widehat{A}^{(2),[0]}_{-+-++-}$ & 2.00000 & -6.00283(9)  & -12.7724(2) & 31.869(6)     & 158.89(6)   \\
  $\widehat{A}^{(2),[1]}_{-+-++-}$ & 0.00000 & -0.583333(1) & 0.96122(6)  & 5.2453(4)     & 2.853(3)    \\
  $\widehat{A}^{(2),[2]}_{-+-++-}$ & 0.00000 & 0.00000      & -0.055555  & -0.240170(1) & -0.25365(2) \\
  \hline
  $q\bar{Q}Q\bar{q}^\prime\bar\nu\ell$  & $\eps^{-4}$ & $\eps^{-3}$ & $\eps^{-2}$ & $\eps^{-1}$ & $\eps^{0}$ \\
  \hline
  $\widehat{A}^{(2),[0]}_{--+++-}$ & 2.00000 & -6.00282(8)  & -18.0013(1)  & 34.592(7)    & 222.52(6)    \\
  $\widehat{A}^{(2),[1]}_{--+++-}$ & 0.00000 & -0.583334(1) &  2.059832(8) & 5.4211(2)    & -13.5049(5)  \\
  $\widehat{A}^{(2),[2]}_{--+++-}$ & 0.00000 & 0.00000      & -0.055555    & -0.081689(2) &  1.10832(2)   \\
  \hline
\end{tabularx}
\end{center}
\caption{The numerical results of
$\hat{A}^{(2),[i]}\left(1_q,2_{\bar{Q}},3_{Q},4_{\bar{q}^\prime},5_{\bar\nu},6_\ell \right)$
using kinematic point in Eq.~\eqref{eq:kinematicpts} for each $(d_s-2)$ component.
}
\label{tab:numQQDs}
\end{table}

\begin{table}[t]
\begin{center}
\begin{tabularx}{0.95\textwidth}{|C{0.9}|C{0.7}|C{1.1}|C{1.2}|C{1.1}|C{1.0}|}
  \hline
  $qgg\bar{q}^\prime\bar\nu\ell$ & $\eps^{-4}$ & $\eps^{-3}$ & $\eps^{-2}$ & $\eps^{-1}$ & $\eps^{0}$ \\
  \hline
  $\widehat{A}^{(2)}_{-++++-}$ & 4.50000 & -3.63577(3) & -277.2182(7) & -344.56(1)  & 2051.1(2) \\
  $          P^{(2)}_{-++++-}$ & 4.5     & -3.63576    & -277.2186    & -344.569(6) & --- \\
  \hline
  $\widehat{A}^{(2)}_{-+-++-}$ & 4.50000 & -3.63581(9) & -13.6826(2) & 6.143(5)   & 66.21(7) \\
  $          P^{(2)}_{-+-++-}$ & 4.5     & -3.63576    & -13.6824    & 6.145(1)   & --- \\
  \hline
  $\widehat{A}^{(2)}_{--+++-}$ & 4.50000 & -3.63579(5) & -18.79219(7) & -6.633(6)  & 79.02(4) \\
  $          P^{(2)}_{--+++-}$ & 4.5     & -3.63576    & -18.79212    & -6.6303(5) & --- \\
  \hline
\end{tabularx}
\end{center}
\caption{The numerical comparison of
$\hat{A}^{(2)}\left(1_q,2_{g},3_{g},4_{\bar{q}^\prime},5_{\bar\nu},6_\ell \right)$
with the universal pole structure $\pole^{(2)}$ defined in Eq.~\eqref{eq:poles}, using kinematic point of
Eq.~\eqref{eq:kinematicpts}, in the HV scheme.}
\label{tab:numggHV}
\end{table}

\begin{table}[t]
\begin{center}
\begin{tabularx}{0.95\textwidth}{|C{0.9}|C{0.7}|C{1.1}|C{1.2}|C{1.1}|C{1.0}|}
  \hline
  $q\bar{Q}Q\bar{q}^\prime\bar\nu\ell$ & $\eps^{-4}$ & $\eps^{-3}$ & $\eps^{-2}$ & $\eps^{-1}$ & $\eps^{0}$ \\
  \hline
  $\widehat{A}^{(2)}_{-+-++-}$ & 2.00000 & -7.16949(9) &  -9.9055(2) & 39.922(6)  & 154.79(7) \\
  $          P^{(2)}_{--+++-}$ & 2       & -7.16944    &  -9.9054    & 39.9245(8) & ---       \\
  \hline
  $\widehat{A}^{(2)}_{--+++-}$ & 2.00000 & -7.16948(8) & -12.9371(1) & 41.432(8)  & 189.53(6) \\
  $          P^{(2)}_{--+++-}$ & 2       & -7.16944    & -12.9370    & 41.4353(6) & --- \\
  \hline
\end{tabularx}
\end{center}
\caption{The numerical comparison of
$\hat{A}^{(2)}\left(1_q,2_{\bar{Q}},3_{Q},4_{\bar{q}^\prime},5_{\bar\nu},6_\ell \right)$
with the universal pole structure $\pole^{(2)}$ defined in Eq.~\eqref{eq:poles}, using kinematic point of
Eq.~\eqref{eq:kinematicpts}, in the HV scheme.}
\label{tab:numQQHV}
\end{table}

We present the results for unrenormalised
$q\bar{Q}Q\bar{q}^\prime\bar\nu\ell$ and $qgg\bar{q}^\prime\bar\nu\ell$
helicity amplitudes, normalised to the tree level amplitude
\begin{equation}
\hat{A}^{(2)}_{\lambda_1\lambda_2\lambda_3\lambda_4\lambda_5\lambda_6}
= \frac{A^{(2)}\left(1^{\lambda_1},2^{\lambda_2},3^{\lambda_3},4^{\lambda_4},5^{\lambda_5},6^{\lambda_6}\right)}
   {A^{(0)}\left(1^{\lambda_1},2^{\lambda_2},3^{\lambda_3},4^{\lambda_4},5^{\lambda_5},6^{\lambda_6}\right)},
\label{eq:treenormalisation}
\end{equation}
with helicities $\lambda_i$. We can further split the amplitude into components of $d_s-2$
\begin{equation}
A^{(2)}\big(1^{\lambda_1},2^{\lambda_2},3^{\lambda_3},4^{\lambda_4},5^{\lambda_5},6^{\lambda_6}\big) =
\sum_{i=0}^{2} \left(d_s-2\right)^i A^{(2),[i]}
\big(1^{\lambda_1},2^{\lambda_2},3^{\lambda_3},4^{\lambda_4},5^{\lambda_5},6^{\lambda_6}\big).
\end{equation}

In Tables~\ref{tab:numggDs}~and~\ref{tab:numQQDs} we display numerical evaluations of the helicity amplitudes for
$qgg\bar{q}^\prime\bar\nu\ell$ and $q\bar{Q}Q\bar{q}^\prime\bar\nu\ell$ channels, respectively, using
the kinematic point given in Eq.~\eqref{eq:kinematicpts} for each $(d_s-2)$ component.
In Tables~\ref{tab:numggHV}~and~\ref{tab:numQQHV},
we compare the divergent part of our numerical results against the
universal two-loop pole structure in Eq.~\eqref{eq:poles} in the HV scheme.
To obtain the two-loop pole structure of Eq.~\eqref{eq:poles} up to the single pole,
we need to compute the one-loop amplitude up to $\cO(\eps)$ for both the
$qgg\bar{q}^\prime\bar\nu\ell$ and $q\bar{Q}Q\bar{q}^\prime\bar\nu\ell$ processes.
The one-loop amplitude is computed by processing the Feynman diagrams through our integrand reduction
setup, followed by IBP decomposition into the one-loop master integral basis consisting of six
bubbles,
a three-mass triangle, two one-mass boxes, a two-mass easy box, two two-mass hard boxes
and a one-mass pentagon, 
in a similar fashion to the two-loop case that is discussed in Section~\ref{sec:setup}. The $\cO(\eps)$ part of the
two-mass easy box and one-mass pentagon integrals are evaluated numerically using
$\textsc{Fiesta}/\textsc{pySecDec}$, while the rest are obtained from available analytic
expressions~\cite{Ellis:2007qk,Papadopoulos:2014lla}. Therefore, the numerical values quoted for the poles in
Tables~\ref{tab:numggHV}~and~\ref{tab:numQQHV} are exact up to $\cO(\eps^{-2})$.
We assess the uncertainty of the $\cO(\eps^{-1})$ part of the pole structure using the same
method as in the two-loop numerical evaluations.

We additionally perform another check of our results by independently
processing the two-loop Feynman diagrams
through a numerical diagram-based integrand reduction into an integrand representation consisting of
four-dimensional ISPs and the extra-dimensional part of the loop momenta $\mu_{ij}$. The integrals
containing $\mu_{ij}$ are first written as dimension-shifted integrals and all integrals appearing
in the amplitude are evaluated directly using $\textsc{Fiesta}/\textsc{pySecDec}$.
This approach is similar to the method we employed for the numerical evaluation of
the planar two-loop five gluon amplitude~\cite{Badger:2017jhb}.
The results obtained using this method are in perfect agreement
with the results reported in this paper.

\section{Conclusions \label{sec:conclusions}}

In this article we have presented numerical results for the planar two-loop helicity amplitudes
for the scattering of a $W$-boson with four partons for the first time. This computation is the first
step towards obtaining analytic expressions using the reconstruction of rational functions
with finite field arithmetic.

A number of important steps remain to be completed in order for this
to be a feasible target. Firstly, the complete list of master integrals should be evaluated
analytically since the sector decomposition approach is still too CPU intensive for phenomenological
applications. This seems a reasonable aim owing to the recent success of the planar and non-planar pentagon
functions~\cite{Chicherin:2017dob,Abreu:2018aqd,Abreu:2019rpt,Chicherin:2018yne,Chicherin:2018mue,Chicherin:2019xeg,Chicherin:2018old}, though the efficient numerical implementation of the resulting analytic functions will require
further study. Secondly, the coefficients of the master integrals still have a high degree of
algebraic complexity. As shown in applications to five-parton scattering, direct reconstruction of
the finite remainder after subtraction of UV and IR poles leads to a substantial reduction in
complexity~\cite{Badger:2018enw,Abreu:2018zmy,Abreu:2019odu}.

We hope that the work presented here presents valuable information that can be used to achieve
these goals, as well as providing encouragement that they are realistic in the near future.

\begin{acknowledgments}
We thank  Johannes Henn, Thomas Gehrmann, Andreas von Manteuffel for helpful discussions.
We also thank Johannes Schlenk for his assistance with \textsc{pySecDec}.
SB is supported by an STFC Rutherford Fellowship ST/L004925/1 and CBH and HBH received partial
support from Rutherford Grant ST/M004104/1. This project has received funding from the European Union’s Horizon
2020 research and innovation programme under grant agreement No 772099. This project
has received funding from the European Union’s Horizon 2020 research and innovation
programme under the Marie Skłodowska-Curie grant agreement 746223.

\end{acknowledgments}

\appendix
\section{Kinematic invariants at the benchmark phase space point}

From the momentum twistor parametrisation in Eq. \eqref{eq:kinematicpts} it is possible to evaluate
all kinematic quantities in the external momenta. Since it may be useful to have reference values
in momentum space we list here six two-particle and three three-particle invariants,
\begin{gather}
s_{12} = -1,\quad
s_{23} = -\frac{83}{102},\quad
s_{34} = -\frac{41584363779551}{5620028969511},\quad
s_{45} = -\frac{12273437608210253843}{7292047345210578060},\nn
\nn
s_{56} = -\frac{137742730207986944}{607670612100881505}, \quad
s_{16} = -\frac{58362131580049744}{321707971112231385}, \quad
s_{123} = -\frac{3226}{2287}, \nn
\nn
s_{234} = -\frac{14812055408288}{9603846973215}, \quad
s_{345} = -\frac{1726859228425207273}{1394067874819669335}.
\end{gather}
In four dimensions only eight of these are independent due to the vanishing of the Gram determinant
of the five independent momenta.

\bibliographystyle{JHEP}
\bibliography{W4pPlanar}

\end{document}

%% file: ExtraDefinitions.tex
\usepackage[utf8]{inputenc}
\usepackage{graphicx}
\usepackage{slashed}
\usepackage{color}
\usepackage{amsmath}
\usepackage{amssymb}

\definecolor{mygreen}{rgb}{0,0.7,0}

\def\cC{\mathcal{C}}

\def\cO{\mathcal{O}}

\def\cA{\mathcal{A}}

\def\nn{\nonumber \\ }

\def\la{\langle}
\def\ra{\rangle}
\def\spA#1#2{\la#1#2\ra}
\def\spB#1#2{[#1#2]}
\def\spAB#1#2#3{\la#1|#2|#3]}

\def\spAA#1#2#3{\la#1|#2|#3\ra}

\def\spaa#1#2#3#4{\la#1|#2|#3|#4\ra}
\def\spbb#1#2#3#4{[#1|#2|#3|#4]}

\DeclareMathOperator{\tr}{\rm tr}
\def\trm{\tr_-}

\def\lrbrace#1{\lbrace#1\rbrace}

\def\eps{\epsilon}

\def\pole{{\mathcal{P}}}
\def\cusp{{\mathrm{cusp}}}

\def\usepix#1#2#3#4#5#6{\parbox{#1}{\includegraphics[width=#1,trim= #3 #4 #5 #6]{#2}}}